\documentclass[twocolumn,showpacs,preprintnumbers,amsmath,amssymb]{revtex4}

\usepackage{graphicx}
\usepackage{dcolumn}
\usepackage{bm}

\begin{document}

\title{Surface spectral function in the superconducting state of a topological insulator}

\author{Lei Hao and T. K. Lee}
 \address{Institute of Physics, Academia Sinica, NanKang, Taipei 11529, Taiwan}

\date{\today}

\begin{abstract}
We discuss the surface spectral function of superconductors
realized from a topological insulator, such as the
copper-intercalated Bi$_{2}$Se$_{3}$. These functions are
calculated by projecting bulk states to the surface for two
different models proposed previously for the topological
insulator. Dependence of the surface spectra on the symmetry of
the bulk pairing order parameter is discussed with particular
emphasis on the odd-parity pairing. Exotic spectra like an
Andreev bound state connected to the topological surface states
are presented.
\end{abstract}


\maketitle

\section{\label{sec:Introduction}Introduction}

Recently, a new state of matter, the topological insulator,
attracts much theoretical and experimental
attention.\cite{kane05,bernevig06,sheng06,konig07,moore07,roy06,fu07l,fu07b,
hsieh08,qi08,xia09,zhang09,hasan10rmp,qi10rev,schnyder08} It
has an electronic structure dominated by the spin-orbit
coupling, which is a band insulator with a well-defined gap in
the bulk but can host an odd number of Dirac cones protected by
time reversal symmetry on the
surface.\cite{kane05,bernevig06,fu07l} The intrinsic spin-orbit
coupling makes it promising for spintronics
applications.\cite{qi08} The proximity induced superconducting
state on the topological insulator surface by a deposited
superconductor was proposed to create Majorana
fermions\cite{wilczek09}, which may provide a new way to
realize the topological quantum computation.\cite{fu0809,sau10}
Lately, the realization of a superconducting state in a typical
topological insulator Bi$_{2}$Se$_{3}$ by intercalating Cu
between adjacent quintuple units (Cu$_{x}$Bi$_{2}$Se$_{3}$)
makes the system even more attractive.\cite{hor10,wray10} The
large diamagnetic response shows that the pairing is mainly of
bulk character. In another topological insulator
Bi$_{2}$Te$_{3}$, the application of a high pressure also turns
the material into a superconducting
state.\cite{zhang10pa,zhang10pb}

Since the superconductivity is bulk and intrinsic to the
material, if zero energy surface Majorana fermion mode exists,
it would be easier to manipulate as compared to that induced by
proximity effect, as it would not experience the interface
roughness or mismatch common to a junction type device.
Possible nontrivial odd-parity pairing in
Cu$_{x}$Bi$_{2}$Se$_{3}$ is proposed and analyzed by Fu and
Berg.\cite{fu10} It was argued that only if a bulk gap opens
and the bulk pairing is odd in parity, would zero energy
Andreev bound states appear in the surface spectrum. However,
their analysis was concentrated on the case when the chemical
potential is much larger than the gap and the topological
surface states are already merged into the continuum conduction
band. A similar analysis was put forward by
Sato.\cite{sato0910}

Despite the above works, a detailed theoretical analysis of
surface spectrum in the superconducting phase arising from a
topological insulator is still lacking. In particular, not much
is reported for the situation when the chemical potential is
only slightly larger than the insulating gap and both
topological surface states (or, the surface conduction
band\cite{wray10}) and the continuum bulk conduction band are
present but separated. Since the surface spectrum is central to
the topological properties of a material both in the normal and
in the superconducting state, which is also directly accessible
by experimental techniques as ARPES\cite{wray10} and
STM\cite{alpichshev10}, it is highly desirable to make a
detailed study of them. This will help to understand better
superconductivity in systems with nontrivial topological band
structure.

We focus on two questions in this paper. One is the effect of
superconducting pairing on the topological surface
states\cite{wray10} present in the normal state. The other is
the existence of surface Andreev bound states. We have noticed
that two different models\cite{zhang09,wang10} are often used
in-discriminatively in literature. They have the same normal
state energy spectra but may be different in the
superconducting state. We thus present our results for both of
them. What happens to the topological surface states when
pairing is introduced in the bulk depends on the orbital
character of the topological surface state and on the bulk
pairing symmetry. Only when the continuum part of the band
opens a full gap and the topological surface states, while
separated from the bulk conduction band, do not open a gap for
an odd-parity pairing, would a gapless Andreev bound state
appear. We find that the orbital characters of the topological
surface states are different for the two models. For a certain
bulk pairing symmetry, it is possible that the topological
surface states of one model opens a gap while that of the other
model is still intact. The existence of Andreev bound states,
the most important indication of nontrivial topological order
in the superconducting phase, is thus expected to be also
related to the orbital character of the topological surface
states. We show that the interplay between the continuum bulk
conduction band and the topological surface states produces a
ring or a segment of zero energy states in addition to the
Andreev bound states depending on the symmetry of bulk pairing
order parameters.

\section{models and the normal state surface modes}

In the following, when we talk about topological insulators, we
would be mainly referring to Bi$_2$Se$_3$, which shows a well
defined Dirac cone structure for the topological surface
states.\cite{zhang09,xia09,hsieh09} The model we consider below
could be easily generalized to study other topological
insulators like TlBiSe$_2$\cite{yan10,kuroda10} and
Bi$_{2}$Te$_{2}$Se\cite{ren10,xu10}.

The band structure of Cu$_{x}$Bi$_{2}$Se$_{3}$ is similar to
that of Bi$_{2}$Se$_{3}$, the most essential part of which
consists of two $p_z$ orbitals on the top and bottom Se layers
hybridized with neighboring Bi $p_z$ orbitals, in each
quintuple Bi$_{2}$Se$_{3}$ unit.\cite{zhang09,fu10} In the
presence of spin-orbit coupling, the normal state has four
degrees of freedom. Label the two orbitals concentrating mainly
on the top and bottom (seeing along the $-z$ direction) Se
layer of the various Bi$_2$Se$_3$ quintuple units as the first
and second orbital, the basis is taken as
$\psi_{\mathbf{k}}$=[$c_{1\mathbf{k}\uparrow}$,
$c_{2\mathbf{k}\uparrow}$, $c_{1\mathbf{k}\downarrow}$,
$c_{2\mathbf{k}\downarrow}$]$^{T}$. The models could be written
compactly in the following matrix
form\cite{zhang09,wang10,fu10}
\begin{equation} \label{h3d0}
H(\mathbf{k})=\epsilon_{0}(\mathbf{k})I_{4\times4}+\sum\limits_{i=0}^{3}m_{i}(\mathbf{k})\Gamma_{i}.
\end{equation}
I$_{4\times4}$ is the fourth order unit matrix, giving rise to
a topologically trivial shift of the energy bands and would be
neglected in most of our following analysis. In terms of the
two by two Pauli matrices $s_{i}$ ($i$=0, $\cdots$, 3) in the
spin subspace and $\sigma_{i}$ ($i$=0, $\cdots$, 3) in the
orbital subspace, the first three Dirac matrices are defined
as\cite{zhang09,wang10} $\Gamma_{0}$=$s_{0}\otimes\sigma_{1}$,
$\Gamma_{1}$=$s_{1}\otimes\sigma_{3}$,
$\Gamma_{2}$=$s_{2}\otimes\sigma_{3}$. As regards $\Gamma_{3}$,
there are presently two different choices:
(I)$s_{0}\otimes\sigma_{2}$\cite{wang10,fu10,li10} and
(II)$s_{3}\otimes\sigma_{3}$\cite{zhang09,liu10b1,liu10b2,lu10,shan10},
which define the two models that would be considered in
parallel in the following discussion. Since Bi$_{2}$Se$_{3}$ is
inversion symmetric, parity could be used to label states. Some
papers adopt the bonding and antibonding states of the two
orbitals defined above as the orbital
basis.\cite{zhang09,li10,liu10b1,liu10b2,lu10,shan10} The
corresponding models could be obtained in terms of a simple
unitary transformation performed in the orbital subspace. Note
that, the two different models give the same bulk band
dispersion
$\epsilon_{\pm}(\mathbf{k})$=$\epsilon_{0}(\mathbf{k})\pm\sqrt{\sum_{i=0}^{3}m_{i}^{2}(\mathbf{k})}$,
with each of two eigen-energies two fold Kramers degenerate due
to the time reversal symmetry and the inversion symmetry of the
model. However, we will show in the following that they are
essentially two physically distinct models.

For the coefficients $\epsilon_{0}(\mathbf{k})$ and
$m_{i}(\mathbf{k})$ ($i=$ 0, 1, 2, 3), there are different
possible parameterizations which coincide with each other close
to the $\Gamma$ point.\cite{zhang09,qi10rev,fu09,wang10}
Without loss of generality, we take the parameterizations of
Wang \emph{et al.}\cite{wang10}. Since the diagonal term
$\epsilon_{0}(\mathbf{k})$ proportional to the unit matrix does
not affect the topological characters, it is ignored here. The
system is defined on a hypothetical bilayer hexagonal lattice
stacked along the $z$-axis, respecting the in plane hexagonal
symmetry of the original Bi$_{2}$Se$_{3}$ lattice. With the
three independent in-plane nearest neighbor unit vectors
defined as $\hat{b}_1$=($\frac{\sqrt{3}}{2}$, $\frac{1}{2}$),
$\hat{b}_2$=(-$\frac{\sqrt{3}}{2}$, $\frac{1}{2}$), and
$\hat{b}_3$=(0, -1), we have
$m_{0}(\mathbf{k})$=$m+2t_{z}(1-\cos
k_{z})+2t(3-2\cos\frac{\sqrt{3}}{2}k_{x}\cos\frac{1}{2}k_{y}-\cos
k_{y})$,
$m_{1}(\mathbf{k})$=2$\sqrt{3}t\sin\frac{\sqrt{3}}{2}k_{x}\cos\frac{1}{2}k_{y}$,
$m_{2}(\mathbf{k})$=2$t(\cos\frac{\sqrt{3}}{2}k_{x}\sin\frac{1}{2}k_{y}+\sin
k_{y})$, and $m_{3}(\mathbf{k})$=$2t_{z}\sin k_{z}$. The
in-plane and out-of-plane lattice parameters\cite{larson02} are
taken as length units in the above expression, that is
$a$=$c$=1. When $mt_z$$<$0 and $mt$$<$0, the parametrization
defined above\cite{wang10} and the parametrization in the small
$\mathbf{k}$ effective model proposed by Zhang \emph{et
al.}\cite{zhang09} describe qualitatively the same physics.
With this parametrization, it is easy to see that the model has
the inversion symmetry $PH(\mathbf{k})P^{-1}$=$H(-\mathbf{k})$,
where the inversion operator is defined as
$P$=$s_0\otimes\sigma_1$.\cite{fu07b}

Now, we clarify the differences between models (I) and (II) by
their surface states, which is one of the most important
signatures of nontrivial topological order in the system. Close
to the $\Gamma$ point in the BZ, we take $m_{i=0, \cdots,
3}(\mathbf{k})$=$\{m+\frac{3}{2}t(k_{x}^2+k_{y}^2)+t_{z}k_{z}^{2},
3tk_{x}, 3tk_{y}, 2t_{z}k_{z}\}$, in which $t>0$, $t_{z}>0$ and
$m<0$. Consider a sample occupying the lower half space
$z\le0$. The possible surface states localized close to $z=0$
is searched by solving a set of four coupled second order
differential equations
\begin{equation}
H(k_{x}=k_{y}=0,k_{z}\rightarrow -i\partial_{z})\Psi(z)=E\Psi(z),
\end{equation}
together with the open boundary condition
$\Psi(z)|_{z=0}$=$\Psi(z)|_{z=-\infty}$=0.\cite{liu10b2,lu10}
$\Psi(z)$ is the four-component eigenvector and E is the energy
of the surface mode, respectively. We look for the zero energy
states and hence set $E$=0.\cite{liu10b2}

For the model (I) with $\Gamma_3$=$s_{0}$$\otimes$$\sigma_{2}$,
the up and down spin degrees of freedom are decoupled from each
other. The wave function could thus be written as
$\Psi(z)$=$[u_{1}(z), u_{2}(z), u_{3}(z),
u_{4}(z)]^{T}$=$[\chi_{\uparrow}(z),
\chi_{\downarrow}(z)]^{T}$. The two spin components of the zero
energy mode satisfy the same equation as ($s$ is $\uparrow$ or
$\downarrow$ for the two spin degrees of freedom)
\begin{equation}
[(m-t_{z}\partial_{z}^{2})\sigma_{1}-2it_{z}\partial_{z}\sigma_{2}] \chi_{s}(z)=0.
\end{equation}
The two degenerate zero energy surface modes for $z$$\le$0 are
obtained as
\begin{equation}
\Psi_{\alpha}(z)=C\eta_{\alpha}(e^{z/\xi_{+}}-e^{z/\xi_{-}}),
\end{equation}
where $\alpha$=1 or 2, $C$ is a normalization constant and
$\xi_{\pm}^{-1}$=$1\pm\sqrt{1+m/t_{z}}$. 1/Re$[\xi_{\pm}^{-1}]$
(`Re' means taking the real part of a number) are the two
penetration depths of the surface modes into the bulk. The two
unit vectors are $(\eta_{1})_{\beta}$=$\delta_{\beta1}$ and
$(\eta_{2})_{\beta}$=$\delta_{\beta3}$, where
$\delta_{\alpha\beta}$ is one for $\alpha$=$\beta$ and zero
otherwise. Take $\{\Psi_{1}, \Psi_{2}\}$ as the two basis, the
effective model for the surface states are obtained by
considering the k$_{x}$ and k$_{y}$ dependent terms in the
original model as perturbations, which are
\begin{equation}
\Delta H_{3D}=\frac{3}{2}t(k_{x}^2+k_{y}^2)\Gamma_{0}+3t(k_{x}\Gamma_{1}+k_{y}\Gamma_{2}).
\end{equation}
Suppose the two basis are normalized, the effective model for
the surface states is\cite{zhang09}
\begin{equation}
H_{eff}(\mathbf{k})=3t(k_{x}s_{x}+k_{y}s_{y}),
\end{equation}
where $s_{x}$ and $s_{y}$ are the first and second Pauli
matrices. Since the two basis both have definite spin
characters, $s_{x}$ and $s_{y}$ in the above equation could
also be considered as acting in the spin subspace. The most
salient feature of this model is that the corresponding surface
states has contributions only from the first orbital. When we
consider a sample occupying $z$$\ge$0, the surface states at
$z$=0 would arise only from the second orbital.

We now study model (II) for
$\Gamma_{3}$=$s_{3}$$\otimes$$\sigma_{3}$. We still consider a
sample situated at $z$$\le$0 with the open boundary conditions.
Following exactly the same steps as for the first model, we
obtain the two degenerate zero energy surface states as
\begin{equation}
\Phi_{\alpha}=C\eta_{\alpha}(e^{z/\xi_{+}}-e^{z/\xi_{-}}),
\end{equation}
where $\alpha$=1 or 2, $C$ is a normalization constant and
$\xi_{\pm}$ are defined identically as above. However, the two
unit basis vectors are quite different from the first model and
are $\eta_{1}$=$\frac{1}{\sqrt{2}}[1, -i, 0, 0]^{T}$ and
$\eta_{2}$=$\frac{1}{\sqrt{2}}[0, 0, -i, 1]^{T}$. The two
dimensional effective model for the surface states is also a
bit different at least formally, which is
\begin{equation}
H_{eff}(\mathbf{k})=3t\hat{z}\cdot(\mathbf{k}\times\mathbf{s})=3t(k_{x}s_{y}-k_{y}s_{x}).
\end{equation}
The Pauli matrices $s_{x}$ and $s_{y}$ act in the two fold
degenerate basis of the zero energy surface states which both
have definite spin characters. For a general two dimensional
wave vector, the surface state would be a linear combination of
all the four spin-orbital basis.

Thus there are qualitative differences between two models which
were used in-discriminatively in the literature for
Bi$_{2}$Se$_{3}$.\cite{wang10,fu10,li10,zhang09,liu10b1,liu10b2,lu10,shan10}
While only one orbital contributes to the surface states for
model (I), both two orbitals contribute in equal weight to the
surface states for model (II). On the other hand, the effective
model of the surface states has the same spin-orbital coupled
form as the linear $k_{x}$ and $k_{y}$ terms in the original
three dimensional model for model (I). However, the effective
model is changed from $\mathbf{k}\cdot\mathbf{s}$ to
$\hat{z}\cdot(\mathbf{k}\times\mathbf{s})$ for model (II). We
have verified that, if we change the in-plane spin-orbit
coupling of the two models (I) and (II) from
$\mathbf{k}\cdot\mathbf{s}$ to
$\hat{z}\cdot(\mathbf{k}\times\mathbf{s})$, the resulting
effective model of the surface states would have the form
$\hat{z}\cdot(\mathbf{k}\times\mathbf{s})$ for model (I) but
will be switched to $\mathbf{k}\cdot\mathbf{s}$ for model (II).

Before ending this section, we would like to point out that
both the $k_{z}$-linear term in $m_{3}(\mathbf{k})$ and the
$k_{z}$-square term in $m_{0}(\mathbf{k})$ are essential to
obtain the zero energy surface modes. If we omit the
$k_{z}^{2}$ term in $m_{0}(\mathbf{k})$, it is easy to verify
that the gapless surface states no longer exist. This is
related to the fact that band inversion is essential to the
appearance of nontrivial topological surface
states\cite{bernevig06,qi10pto}, which can only occur in the
presence of $k_{z}^{2}$ term for $k_{x}$=$k_{y}$=0.

\section{superconducting state spectral function}

\subsection{surface Green's functions in the superconducting state}

The realization of the superconducting state in
Cu$_{x}$Bi$_{2}$Se$_{3}$\cite{hor10,wray10} has brought about
excitement that non-trivial topological superconducting state
might be realized in this system, in which topologically
protected gapless surface states traverse the bulk
superconducting gap.\cite{fu10,sato0910,qi10b} The recent
realization of superconducting phase in Bi$_{2}$Te$_{3}$ under
high pressure\cite{zhang10pa,zhang10pb} makes the
Bi$_{2}$X$_{3}$ (X is Se or Te) material a very promising
candidate system to realize topologically nontrivial
superconducting phases.\cite{sato0910,fu10,qi10b}

The normal state of the topological insulator is marked by the
presence of topological surface states inside the bulk gap.
These gapless topological surface states are well separated
from the bulk conduction band at low energies and become
indistinguishable for energies much higher than the conduction
band minimum. Depending on the doped charge density the
superconducting state could occur with the chemical potential
either deep in the bulk conduction band or in the intermediate
region where the topological surface states are well separated
from the bulk conduction band\cite{wray10}. In the latter case,
the coupling between the continuum bulk states and the isolated
topological surface states may cause some new interesting
phenomena. Thus this intermediate region is where we will
concentrate on below. Furthermore, the actual pairing
symmetries of the superconducting Cu$_{x}$Bi$_{2}$Se$_{3}$ and
Bi$_{2}$Te$_{3}$ are presently
unknown.\cite{hor10,wray10,zhang10pa,zhang10pb} Thus we will
examine cases with different pairing symmetries in the hope to
provide clues to identify the pairing symmetry and the role
contributed by the topological surface states.

In the following, we will study the surface spectral function
to see possible nontrivial topological properties arising from
the normal phase topological order which is subject to a
certain bulk pairing. The surface spectral function, which
could be obtained from the surface Green's function, has been
studied by ARPES\cite{wray10} and STM\cite{alpichshev10} to
give important information on the topological properties of the
system. In the superconducting state, we expect to see some
surface Andreev bound states if a certain superconducting order
is realized in the material.

In the presence of a surface perpendicular to the $z$-axis,
$k_{x}$ and $k_{y}$ are good quantum numbers, and $k_{z}$ is
replaced by $-i\partial_{z}$ as we shall search for surface
states. We then discretize the $z$ coordinate and turn the
whole sample ($z$$\le$0) with a surface at $z$=0 to a coupled
quintuple-layer system. Label each separate quintuple unit with
an integer index $n$, and make the substitutions
$\partial_{z}\psi_{n}(z)$=$\frac{1}{2}[\psi_{n+1}-\psi_{n-1}]$
and
$\partial_{z}^{2}\psi_{n}(z)$=$\psi_{n+1}+\psi_{n-1}-2\psi_{n}$
($c$ is set as length unit along $z$ axis), the Hamiltonian
consists now the intra-layer terms and the interlayer hopping
terms, $\hat{H}$=$\hat{H}_{\parallel}+\hat{H}_{\perp}$. The
intra-layer part of the model is
$\hat{H}_{\parallel}$=$\sum_{n\mathbf{k}}
\psi_{n\mathbf{k}}^{\dagger}h_{xy}(\mathbf{k})\psi_{n\mathbf{k}}$,
in which
\begin{equation}
h_{xy}(\mathbf{k})=m'_{0}(\mathbf{k})\Gamma_{0}+m_{1}(\mathbf{k})\Gamma_{1}
+m_{2}(\mathbf{k})\Gamma_{2}.
\end{equation}
$m_{1}(\mathbf{k})$ and $m_{2}(\mathbf{k})$ are the same as
those in the bulk model. $m'_{0}(\mathbf{k})$ is obtained from
$m_{0}(\mathbf{k})$ by first expanding it up to the square term
of $k_{z}$ and then replacing the term proportional to
$k_{z}^{2}$ from $Ck_{z}^{2}$ to $2C$.

The inter-layer hopping term is
$\hat{H}_{\perp}$=$\sum_{n\mathbf{k}}\psi_{n\mathbf{k}}^{\dagger}h_{z}\psi_{n+1,\mathbf{k}}$+H.c.
In terms of the parameterizations of Wang \emph{et
al.}\cite{wang10}, we have
\begin{equation}
h_{z}=-t_{z}(\Gamma_{0}+i\Gamma_{3}).
\end{equation}
Since the $\Gamma_{3}$ matrix now appears only in the $h_{z}$
part of the coupled layers system, the difference between the
two models enters only through the interlayer hopping term.

Now introduce superconducting pairing and define the Nambu
basis as
$\phi_{n\mathbf{k}}^{\dagger}$=$[\psi_{n\mathbf{k}}^{\dagger},
\psi_{n-\mathbf{k}}^{T}]$. The intra-layer part of the
Bogoliubov de Gennes (BdG) Hamiltonian is then
$\hat{H}^{SC}_{\parallel}$=$\sum_{n\mathbf{k}}\phi_{n\mathbf{k}}^{\dagger}H_{SC}(\mathbf{k})\phi_{n\mathbf{k}}$,
in which\cite{linder10}
\begin{equation}
H_{SC}(\mathbf{k})=
\begin{pmatrix} h_{0}(\mathbf{k}) & \underline{\Delta}(\mathbf{k}) \\
-\underline{\Delta}^{\ast}(-\mathbf{k}) & -h^{\ast}_{0}(-\mathbf{k})
\end{pmatrix},
\end{equation}
where $h_{0}(\mathbf{k})$=$h_{xy}(\mathbf{k})-\mu
I_{4\times4}$, with $\mu$ the chemical potential.
$\underline{\Delta}(\mathbf{k})$ is the 4$\times$4 pairing
matrix. Ignoring the possibility of interlayer pairing, the
interlayer hopping terms are
$\hat{H}^{SC}_{\perp}$=$\sum_{n\mathbf{k}}\phi_{n\mathbf{k}}^{\dagger}H_{z}\phi_{n+1,\mathbf{k}}$+H.c.,
in which
\begin{equation}
H_{z}=
\begin{pmatrix} h_{z} & 0 \\
0 & -h^{\ast}_{z}
\end{pmatrix}.
\end{equation}

Once the pairing order is given, the surface spectral function
is obtained from the retarded surface Green's functions, which
could be calculated in terms of standard transfer matrix
method.\cite{fastiter} In the simplest form of the method, the
8$\times$8 retarded surface Green's function
$G(\mathbf{k},\omega)$ is obtained by self-consistent
calculation of $G(\mathbf{k},\omega)$ and a transfer matrix
$T(\mathbf{k},\omega)$ as\cite{fastiter,wang10}
\begin{subequations}
\begin{equation}
G^{-1}=g^{-1}-H_{z}^{\dagger}T,
\end{equation}
\begin{equation}
T=GH_{z},
\end{equation}
\end{subequations}
where $g$=$[zI_{8\times8}-H_{SC}(\mathbf{k})]^{-1}$
($z$=$\omega+i\eta$) is the retarded Green's function for an
isolated layer. $\eta$ is the positive infinitesimal 0$^{+}$,
which is replaced by a small positive number in realistic
calculations. Self-consistent calculation of the Green's
function starts with $G$=$g$. The surface Green's function
could also be obtained in terms of other iteration schemes,
such as the algorithm in Ref.\cite{fastiter}. We have found no
difference between the results obtained in terms of different
iteration schemes.

After the retarded Green's functions are at hand, the spectral
function is obtained as
\begin{equation}
A(\mathbf{k},\omega)=-\sum_{i=1}^{4}\text{Im} G_{ii}(\mathbf{k},\omega)/\pi.
\end{equation}

Since we have now two orbital and two spin degrees of freedom,
there are many possible pairing channels for different possible
pairing mechanisms. Realistic theoretical determination of the
pairing symmetry requires the knowledge of pairing mechanism
and reasonable parameter values, which are both lacking
presently.\cite{fu10} Here we will consider singlet and triplet
pairing orders as phenomenological input parameters. Their
qualitative differences in spectral functions could help to
identify the pairing symmetry from experiments.

\subsection{gap opening in the topological surface states}

Before presenting the full spectral function, we first would
like to examine what happens to the topological surface states
inherited from the normal state\cite{wray10} upon the formation
of a certain bulk pairing. The most salient feature of the
topological insulator is the presence of gapless surface states
(3D) or edge states (2D).\cite{kane05,bernevig06,fu07l,qi10pto}
In the case of Cu$_{x}$Bi$_{2}$Se$_{3}$, it is found that these
surface states in the non-superconducting Bi$_{2}$Se$_{3}$
persist to the superconducting copper intercalated samples and
are well separated from the bulk conduction band and hence
well-defined.\cite{hor10,wray10} It is thus an interesting
question what would happen to them if a certain pairing forms
in the bulk. In this subsection, we give a simple criterion to
judge whether a gap would be induced in the topological surface
states for an arbitrary bulk pairing.

Suppose the chemical potential lies slightly above the bottom
of the bulk conduction band where the topological surface
states are well separated and well defined.\cite{wray10} If a
pairing is realized in the topological surface states, it
should occur between the two time reversal related states for
$\mathbf{k}$ and -$\mathbf{k}$.\cite{fu0809}

We first consider model (I). For our purpose, we would
concentrate on the positive energy branch of the topological
surface states. When the pairing occurs in the valence
band\cite{zhang10pa,zhang10pb}, the analysis and conclusion
would be similar. Since pairing occurs in the (k$_{x}$,
k$_{y}$) space, we would ignore the $z$-dependence of the
surface modes when analyzing pairing properties. From the basis
and the effective model obtained in Sec. II, the two
eigenvectors for a certain 2D wave vector are
\begin{equation}
\eta_{\alpha}(\mathbf{k})=\frac{1}{\sqrt{2}}[1, 0, \alpha\frac{k_{+}}{k}, 0]^{T},
\end{equation}
where $\alpha$ is `+' (`$-$') for the upper (lower) branch of
the surface states, $k_{\pm}$=$k_{x}\pm ik_{y}$,
$k$=$\sqrt{k_{x}^2+k_{y}^2}$. The annihilation operators of
these states are
\begin{equation}
d_{\mathbf{k}\alpha}=\frac{1}{\sqrt{2}}[c_{1\mathbf{k}\uparrow}
+\frac{\alpha k_{+}}{k}c_{1\mathbf{k}\downarrow}]
\end{equation}
If pairing is induced in the upper surface conduction band at
$\mathbf{k}$, the only possible pairing would be proportional
to
$d_{\mathbf{k}+}^{\dagger}d_{-\mathbf{k}+}^{\dagger}$.\cite{fu0809}
Denote the time-reversal operator as
$\mathcal{T}$.\cite{fu0809,qi10b} Since
$\mathcal{T}c_{1\mathbf{k}\uparrow}^{\dagger}\mathcal{T}^{-1}$=$c_{1-\mathbf{k}\downarrow}^{\dagger}$
and
$\mathcal{T}c_{1\mathbf{k}\downarrow}^{\dagger}\mathcal{T}^{-1}$=$-c_{1-\mathbf{k}\uparrow}^{\dagger}$,
we have $\mathcal{T}
d_{\mathbf{k}+}^{\dagger}d_{-\mathbf{k}+}^{\dagger}\mathcal{T}^{-1}$
=$\frac{k_{+}^{2}}{k^2}d_{\mathbf{k}+}^{\dagger}d_{-\mathbf{k}+}^{\dagger}$.
To ensure the time-reversal symmetry of the pairing, the actual
pairing should be of the form
\begin{eqnarray}
\hat{\Delta}^{I}_{SCB}(\mathbf{k})&=&\Delta_{0}\frac{k_{+}}{k}
d_{\mathbf{k}+}^{\dagger}d_{-\mathbf{k}+}^{\dagger} \notag \\
&=&\frac{\Delta_{0}}{2}[\frac{k_{+}}{k}c_{1\mathbf{k}\uparrow}^{\dagger}c_{1-\mathbf{k}\uparrow}^{\dagger}
-\frac{k_{-}}{k}c_{1\mathbf{k}\downarrow}^{\dagger}c_{1-\mathbf{k}\downarrow}^{\dagger} \notag \\
&&+(c_{1\mathbf{k}\downarrow}^{\dagger}c_{1-\mathbf{k}\uparrow}^{\dagger}
-c_{1\mathbf{k}\uparrow}^{\dagger}c_{1-\mathbf{k}\downarrow}^{\dagger})],
\end{eqnarray}
where $\Delta_{0}$ is the real pairing amplitude, which could
be an even or odd real function of $\mathbf{k}$ depending on
the pairing realized in the bulk. `SCB' is abbreviation for the
surface conduction band (the topological surface states). Thus,
the surface conduction band only supports the anti-phase
$p_{x}\pm ip_{y}$ equal-spin triplet pairing and the
spin-singlet pairing within orbital 1. No other bulk pairing
channels, especially those inter-orbital pairings, would open a
gap in the topological surface states within the framework of
model (I).

For model (II), the two eigenvectors of the surface states for
a certain 2D wave vector are (again, ignoring the
$z$-dependence)
\begin{equation}
\eta_{\alpha}(\mathbf{k})=\frac{1}{2}[1, -i, \alpha\frac{k_{+}}{k}, i\alpha\frac{k_{+}}{k}]^{T},
\end{equation}
where $\alpha$ is `+' (`$-$') for the upper (lower) branch of
the topological surface states. Following the same arguments as
for the first model,  when the chemical potential cuts the
upper branch of these well defined surface states the time
reversal invariant pairing is of the form
\begin{eqnarray}
\hat{\Delta}^{II}_{SCB}(\mathbf{k})&=&\Delta_{0}\frac{k_{+}}{k}
d_{\mathbf{k}+}^{\dagger}d_{-\mathbf{k}+}^{\dagger} \notag \\
&=&\frac{\Delta_{0}}{4}[\frac{k_{+}}{k}(c_{1\mathbf{k}\uparrow}^{\dagger}c_{1-\mathbf{k}\uparrow}^{\dagger}
-c_{2\mathbf{k}\uparrow}^{\dagger}c_{2-\mathbf{k}\uparrow}^{\dagger})  \notag \\
&&-\frac{k_{-}}{k}(c_{1\mathbf{k}\downarrow}^{\dagger}c_{1-\mathbf{k}\downarrow}^{\dagger}
-c_{2\mathbf{k}\downarrow}^{\dagger}c_{2-\mathbf{k}\downarrow}^{\dagger}) \notag \\
&& +i\frac{k_{+}}{k}(c_{1\mathbf{k}\uparrow}^{\dagger}c_{2-\mathbf{k}\uparrow}^{\dagger}
+c_{2\mathbf{k}\uparrow}^{\dagger}c_{1-\mathbf{k}\uparrow}^{\dagger})  \notag \\
&& +i\frac{k_{-}}{k}(c_{1\mathbf{k}\downarrow}^{\dagger}c_{2-\mathbf{k}\downarrow}^{\dagger}
+c_{2\mathbf{k}\downarrow}^{\dagger}c_{1-\mathbf{k}\downarrow}^{\dagger})  \notag \\
&& +(c_{1\mathbf{k}\downarrow}^{\dagger}c_{1-\mathbf{k}\uparrow}^{\dagger}
-c_{1\mathbf{k}\uparrow}^{\dagger}c_{1-\mathbf{k}\downarrow}^{\dagger}  \notag \\
&& +c_{2\mathbf{k}\downarrow}^{\dagger}c_{2-\mathbf{k}\uparrow}^{\dagger}
-c_{2\mathbf{k}\uparrow}^{\dagger}c_{2-\mathbf{k}\downarrow}^{\dagger})  \notag \\
&& +i(c_{1\mathbf{k}\uparrow}^{\dagger}c_{2-\mathbf{k}\downarrow}^{\dagger}
+c_{1\mathbf{k}\downarrow}^{\dagger}c_{2-\mathbf{k}\uparrow}^{\dagger}  \notag \\
&& -c_{2\mathbf{k}\uparrow}^{\dagger}c_{1-\mathbf{k}\downarrow}^{\dagger}
-c_{2\mathbf{k}\downarrow}^{\dagger}c_{1-\mathbf{k}\uparrow}^{\dagger})].
\end{eqnarray}
As in Eq. (17), $\Delta_{0}$ could be a constant or a real
function of $\mathbf{k}$ compatible with the symmetry of one
pairing component contained in the above decomposition. Besides
the intra-orbital pairing channels active in the model (I),
there are two additional inter-orbital pairing channels that
are effective in producing a gap in the topological surface
states. The last term in the above equation is just the
odd-parity inter-orbital triplet pairing proposed by Fu and
Berg\cite{fu10} as a possible candidate of a topological
superconductor to be realized in a superconductor like
Cu$_{x}$Bi$_{2}$Se$_{3}$.

The clear difference between $\hat{\Delta}^{I}_{SCB}$ and
$\hat{\Delta}^{II}_{SCB}$ makes the distinction between model
(I) and model (II) more obvious. Since the gap opening of the
surface states is measurable, it is highly desirable to
ascertain which model is the correct description of the
underlying physics of Bi$_{2}$Se$_{3}$ and Bi$_{2}$Te$_{3}$.

Previously, a simple effective model calculation indicates that
no gap opens in the topological surface states for any triplet
pairing induced by proximity effect on the surface of a
topological insulator.\cite{linder10} However, our analysis
above indicates that if the proximity induced triplet pairing
is compatible with any of the triplet components explicit in
$\hat{\Delta}^{I}_{SCB}$ ($\hat{\Delta}^{II}_{SCB}$) for model
I (model II), then a full pairing gap could still be opened in
the topological surface states. Note that the real gap opening
pattern in the topological surface states also depends on
$\Delta_0$.

Except for the pairing channels explicit in
$\hat{\Delta}^{I}_{SCB}$ for model (I) and
$\hat{\Delta}^{II}_{SCB}$ for model (II), no other bulk pairing
could open a gap in the topological surface states. The
existence of surface Andreev bound states depends on whether or
not a gap opens in the topological surface states. We clarify
this matter in the next section.

\subsection{spectral function for typical pairing symmetries}

Observation of superconductivity in Cu$_{x}$Bi$_{2}$Se$_{3}$
brings about anticipation that nontrivial topological
superconducting states might be realized in this material. The
topological superconductor is defined as a state with a full
pairing gap in the bulk and nontrivial gapless Andreev bound
states on the surface.\cite{fu10}

Possible pairings realizable in a system depend on the symmetry
of the system and the specific pairing mechanism. In the case
of pairing induced by short range electron density-density
interactions, Fu and Berg identified four possible pairing
channels.\cite{fu10} However, if the pairing is induced by more
long-range interactions, such as the electron-phonon
interaction, other pairing channels (e.g., in which the pairing
potential is $\mathbf{k}$ dependent) would also be possible. In
the following we would analyze several typical pairings and
compare results of the two different models. In each case,
there are three typical situations as regards to the position
of the chemical potential $\mu$: (1) $\mu$ lies in the bulk
gap; (2) $\mu$ lies above but close to the bottom of the bulk
conduction band, where the topological surface states are well
separated from the continuum bulk conduction band; (3) $\mu$
lies far above the bulk conduction band bottom, where the
surface states have merged into the continuum conduction band.
While the latter two cases are relevant to the superconducting
state of Cu$_{x}$Bi$_{2}$Se$_{3}$\cite{hor10,wray10}, the first
case could be regarded as mimicking the proximity effect from
an external superconductor.\cite{fu0809,sau10,linder10} In this
paper we would focus on the latter two situations. When the
chemical potential lies in the valence
band\cite{zhang10pa,zhang10pb}, the results should be
qualitatively similar for the same type of bulk pairing.

Follow Wang \emph{et al.}\cite{wang10}, the model parameters
are taken as $t$=$t_{z}$=0.5, $m$=-0.7 in most cases. 0.7 is
half of the bulk band gap. The width of the bulk conduction
band at $k_{x}$=$k_{y}$=0 is $2(2t_{z}-|m|)$=0.6. The small
positive number $\eta$ in the Green's functions is taken as
10$^{-4}$.

\emph{- even-parity intra-orbital singlet pairing} First, we
study the simplest possible pairing denoted by
$\underline{\Delta}(\mathbf{k})$=$i\Delta_{0}s_{2}\otimes\sigma_{0}$.
Spectral functions for the two different models are the same
for this pairing, so only one is presented in Fig. 1. Here and
in the following, the degree of darkness indicates the
intensity of the spectrum. The continuum portions of spectrum
are contributions from the bulk states, which have small finite
amplitudes on the surface. Henceforth, they would be called
bulk conduction band for simplicity. The contributions from the
topological surface states are somewhat speckled because we
have taken a finite grid in the $(\mathbf{k},\omega)$ plane to
calculate the spectral function. When the grid points are taken
to be very dense, contributions from the topological surface
states will also become smooth. To see the qualitative behavior
more clearly, a reasonably large pairing amplitude
$\Delta_{0}$=0.1 is considered.\cite{wray10} The result is
nearly identical in the $\Gamma$K direction (along $k_{x}$
axis) and the $\Gamma$M direction (along $k_{y}$ axis) of the
2D reduced Brillouin zone (BZ). The topological surface states
of both two models open a gap, which are consistent with the
analysis of the previous subsection. Since no Andreev bound
state exists, this pairing is topologically trivial. The other
intra-orbital singlet pairings with a $\mathbf{k}$-dependent
$\Delta_{0}$, which is an even function of $\mathbf{k}$ could
also be considered, such as the $d_{x^2-y^2}$-wave pairing. In
these cases, there would be line nodes along the nodal
directions of the pairing gap.

\begin{figure}
\centering
\includegraphics[width=9cm,height=10cm,angle=0]{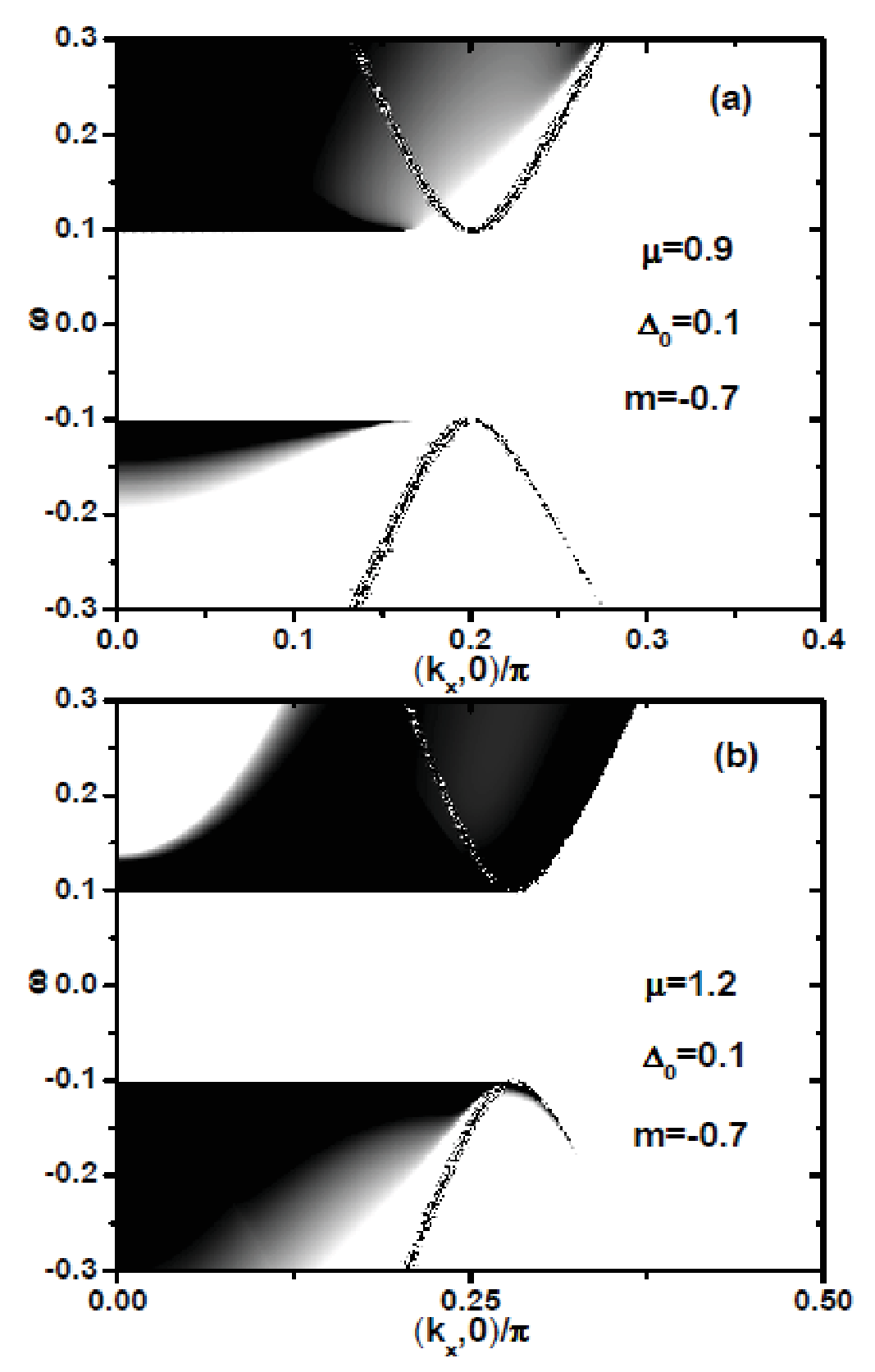}
\caption{Spectral function for even-parity intra-orbital $s$-wave pairing,
for two typical parameter sets for which the topological surface states at the chemical potential
(a) are well separated from the bulk conduction band
and (b) merges into the bulk conduction band.
The two models give identical results for this pairing. Spectrum along other directions
(crossing the $\Gamma$ point) are qualitatively identical.}
\end{figure}

\emph{- even-parity inter-orbital singlet pairing} Since there
are now two orbits, another singlet pairing exists in the
inter-orbital channel. The pairing matrix for the $s$-wave case
is
$\underline{\Delta}(\mathbf{k})$=$i\Delta_{0}s_{2}\otimes\sigma_{1}$.
The corresponding spectral functions presented in Fig. 2 for
this pairing are still identical for the two models. They
differ from the spectra of the former intra-orbital pairing
channel in at least two aspects. First, no gap opens in the
topological surface states, which is in agreement with the
criterion proposed in the previous subsection. Second, though a
full gap also opens in the continuum part of the spectrum, it
is not constant and shows some $\mathbf{k}$ dependence. As
shown in Fig. 2(b), the continuum part of the spectrum even
nearly closes at some special wave vectors for certain
parameters. Another interesting feature is the strong
redistribution of spectral weight between the continuum
conduction band and the topological surface states. Some weight
in the bulk conduction band part of the surface spectrum above
the chemical potential is depleted and transferred to the
topological surface states below the chemical potential. This
redistribution arises from the particle hole mixing induced by
the presence of bulk superconducting pairing. As would see
below, a similar feature is present for each bulk pairing that
does not open a gap in the topological surface states.

\begin{figure}
\centering
\includegraphics[width=9cm,height=15cm,angle=0]{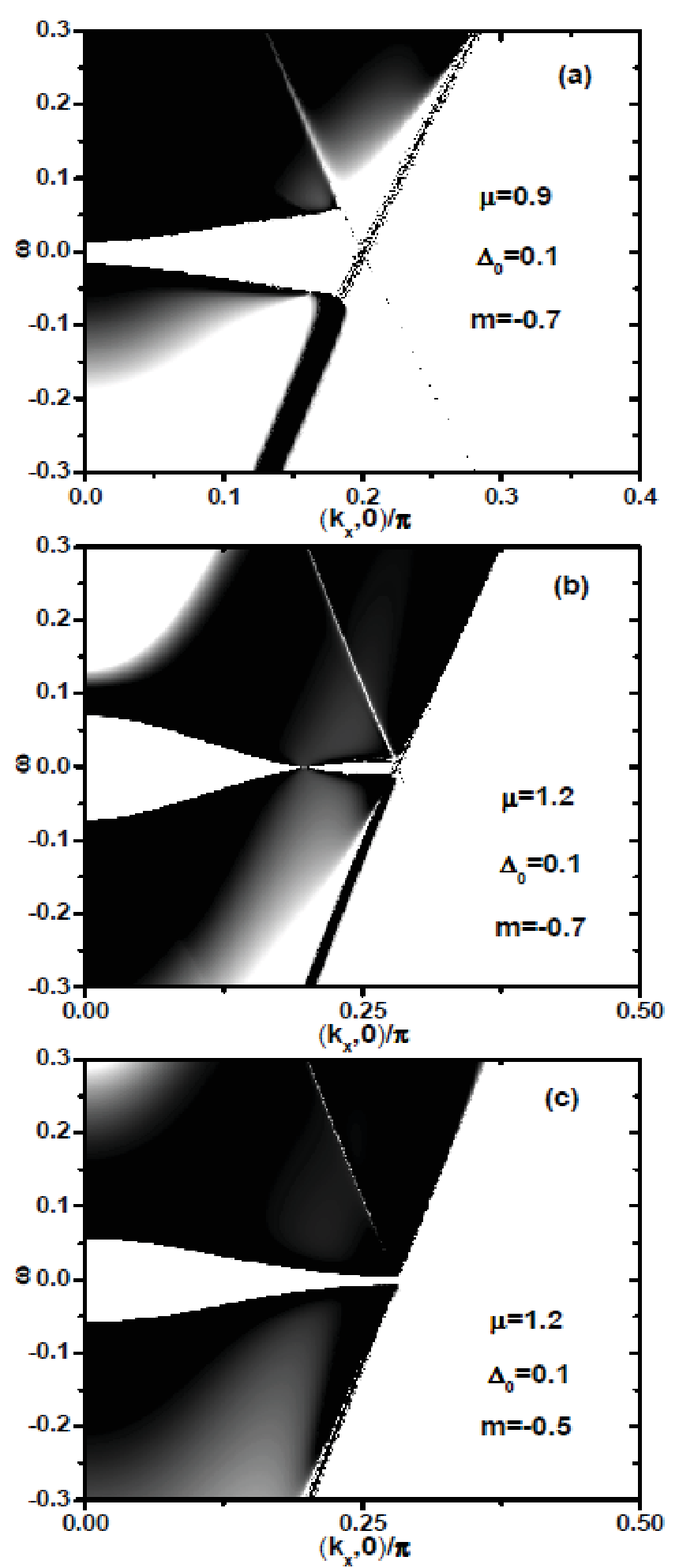}
\caption{Spectral function for even-parity inter-orbital $s$-wave pairing,
for three sets of parameters for which at the chemical potential
(a) the topological surface states are well separated from the bulk conduction band,
(b) the topological surface states are almost merged into the bulk conduction band,
and (c) the topological surface states are well merged into the bulk conduction band.
The two models give identical results for this pairing. Spectrum along other directions
(crossing the $\Gamma$ point) are qualitatively identical.}
\end{figure}

\emph{- odd-parity inter-orbital triplet pairing} We now
consider the odd-parity inter-orbital triplet pairing channel,
proposed by Fu and Berg as a candidate for possible nontrivial
topological superconducting states in
Cu$_{x}$Bi$_{2}$Se$_{3}$.\cite{fu10} The pairing matrix is
$\underline{\Delta}(\mathbf{k})$=$\Delta_{0}s_{1}\otimes\sigma_{2}$.
As was shown in Fig. 3, the spectral functions for the two
models differ greatly. When the chemical potential is close to
bottom of the bulk conduction band, the surface conduction band
is still gapless for model (I) but opens a gap for model (II),
in agreement with the analysis in the above subsection. Another
essential difference is the existence or not of Andreev bound
states. For model (I), a band of Andreev bound states appears
inside of the insulating gap of the continuum which
continuously connects to the topological surface states. While
for model (II), there is a point node at (0, 0), no Andreev
bound state exists inside of the gap region. When the chemical
potential is increased to the position where the surface
conduction band has almost merged into the continuum part of
the surface spectrum (corresponding to contribution from the
bulk conduction band), surface spectra for the two models are
as shown in Fig. 4. Since now there is no well separated
surface conduction band, a full gap opens also for the model
(I). However, a band of Andreev bound states still exists. When
we further increase the chemical potential to $\mu>1.3$ (for
$m$=-0.7), there is no state close to the $\Gamma$ point, then
there would be no Andreev bound state even for model (I). A
strong redistribution of spectral weight arising from the
particle hole mixing between the continuum bulk conduction band
and the topological surface states is observed in results for
model (I).

\begin{figure}
\centering
\includegraphics[width=9cm,height=10cm,angle=0]{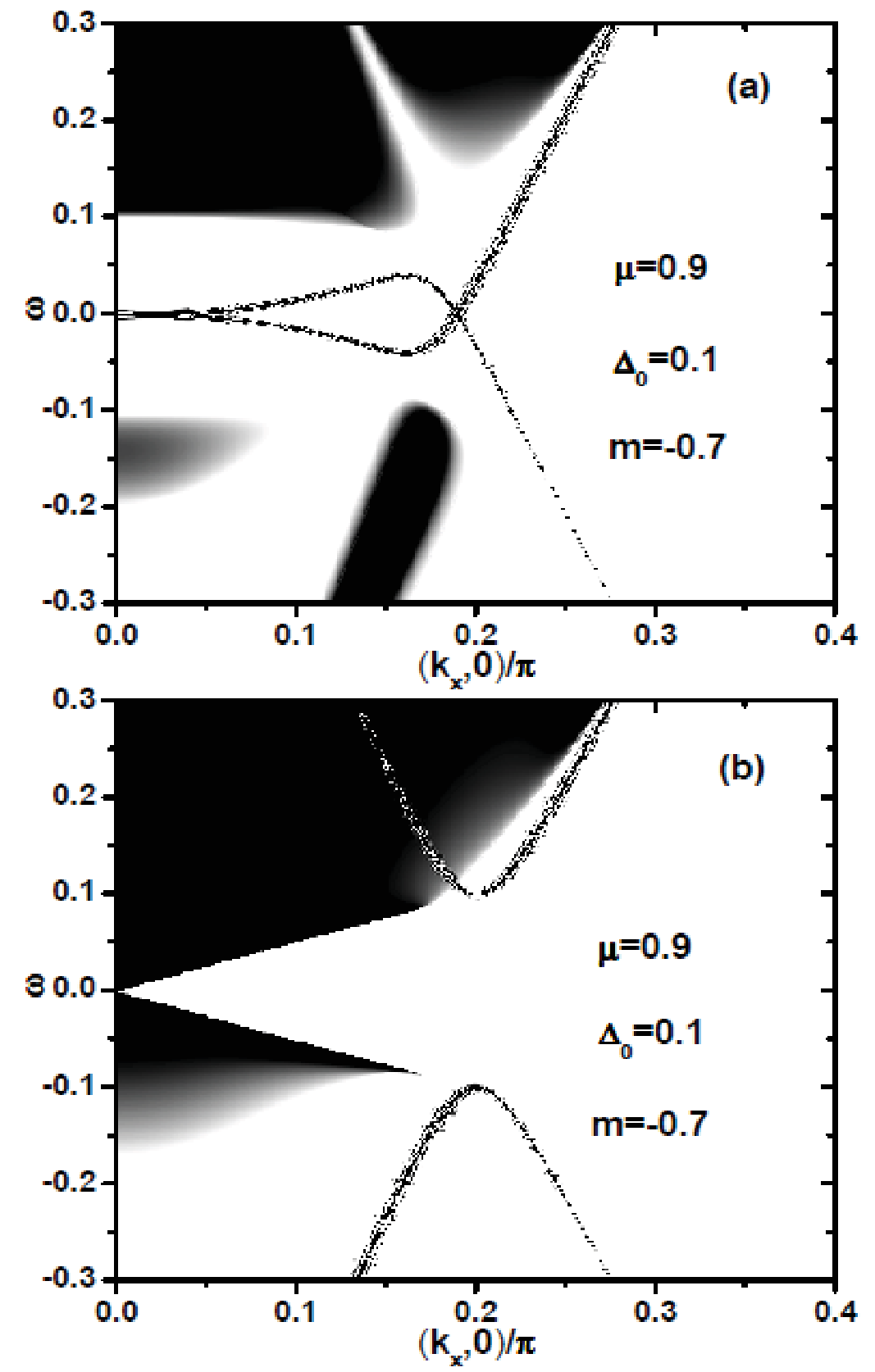}
\caption{Spectral function for odd-parity inter-orbital triplet pairing
in the opposite spin pairing channel, for
(a) model (I) and (b) model (II). As shown are the cases for which
the topological surface states are well separated from the bulk conduction band at the chemical potential.
The parameters are as shown on the figures. Spectrum along other directions
(crossing the $\Gamma$ point) are qualitatively identical.}
\end{figure}

\begin{figure}
\centering
\includegraphics[width=9cm,height=10cm,angle=0]{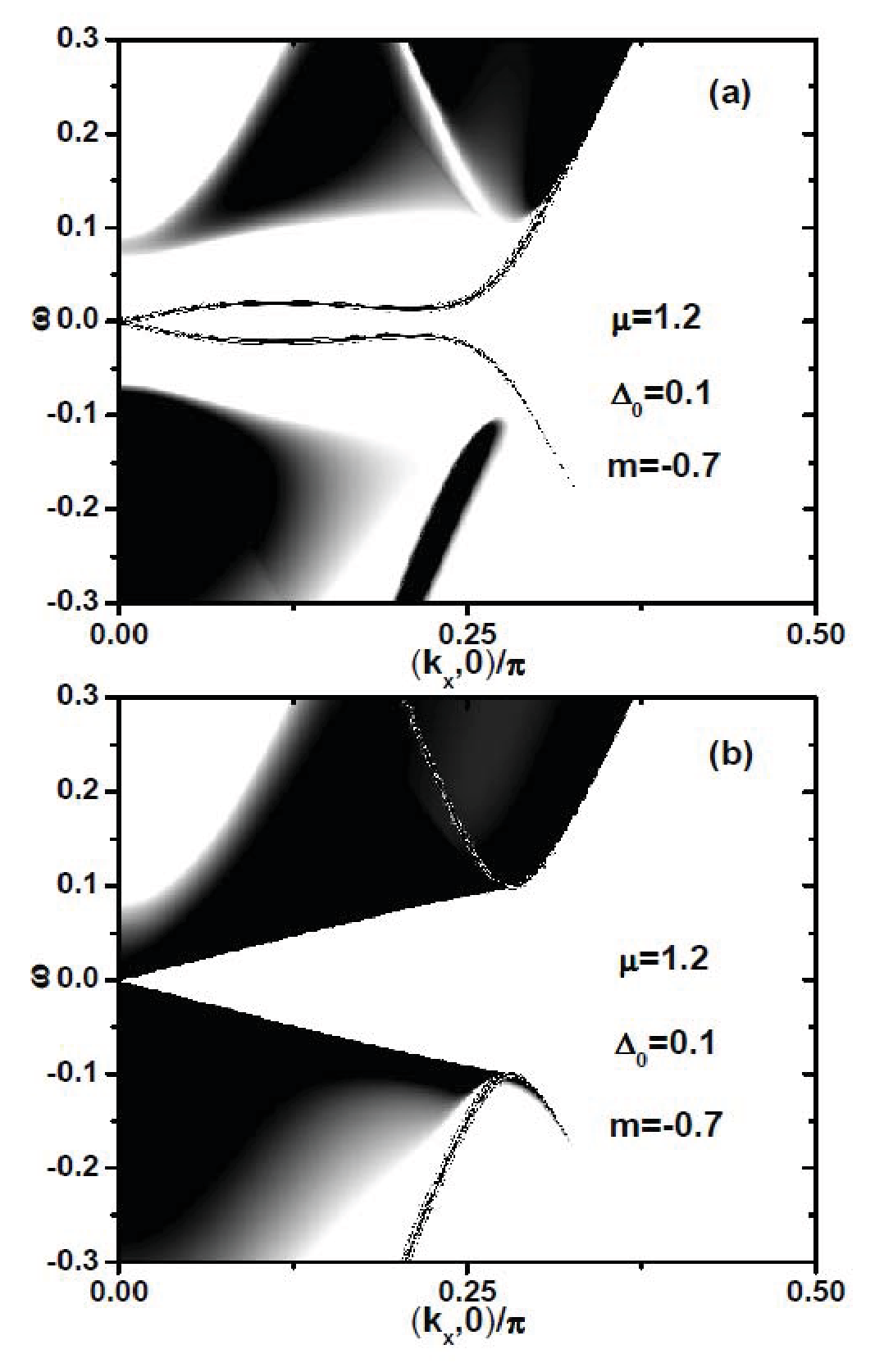}
\caption{Spectral function for odd-parity inter-orbital triplet pairing
in the opposite spin pairing channel, for
(a) model (I) and (b) model (II). As shown are the cases for which
the topological surface states are merged into the bulk conduction band at the chemical potential.
The parameters are as shown on the figures. Spectrum along other directions
(crossing the $\Gamma$ point) are qualitatively identical.}
\end{figure}

Besides the pairing studied above, there are also other
odd-parity pairings. Since they are possibly related with
nontrivial topological superconducting phases, we analyze
several of them in the following.

\emph{- odd-parity intra-orbital singlet pairing} In this case,
every orbit pairs into a spin-singlet, but the two orbits has a
relative $\pi$ phase difference and is hence odd in
parity.\cite{fu10} The pairing matrix is denoted as
$\underline{\Delta}(\mathbf{k})$=$i\Delta_{0}s_{2}\otimes\sigma_{3}$.
The corresponding spectral functions for the two models are
presented in Figs. 5(a) and 5(b) for chemical potential close
to the bottom of the bulk conduction band and thus the
topological surface state
 is well defined. A comparison with the previous
pairing, shown in Figs. 3 and 4, indicates that the results are
interchanged between the two models. Gap opening in the
topological surface states again follows the expectation from
the previous subsection. The Andreev bound states in the bulk
band gap for the second model again connect continuously to the
protected topological surface states. The behaviors for higher
chemical potentials, as shown in Figs. 5(c) and 5(d), are
similar to that of the previous pairing shown in Fig. 4 with
the two models interchanged.

\begin{figure}
\centering
\includegraphics[width=9cm,height=8cm,angle=0]{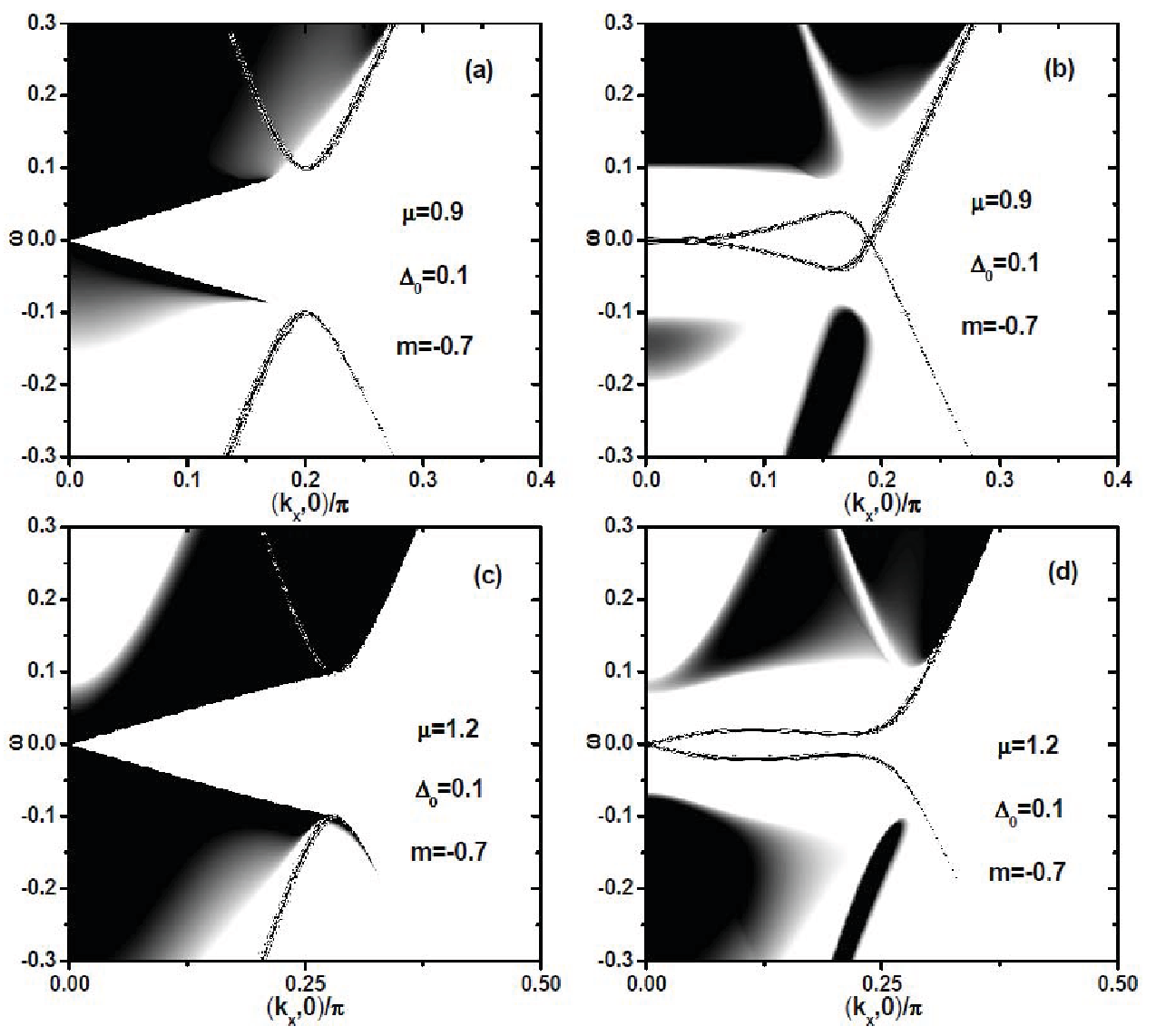}
\caption{Spectral function for odd-parity intra-orbital $s$-wave pairing, for
model (I) ((a) and (c)) and model (II) ((b) and (d)).
For (a) and (b) ((c) and (d)), the topological surface states
are well separated from (merged into) the bulk conduction band
at the chemical potential. Spectrum along other directions
(crossing the $\Gamma$ point) are qualitatively identical.}
\end{figure}

\emph{- odd-parity inter-orbital equal-spin triplet pairing}
Another interesting possibility is the equal-spin pairing
channel. Here we consider the two fold degenerate inter-orbital
pairing channels as proposed by Fu and Berg.\cite{fu10} This
kind of pairing could be favored by interorbital ferromagnetic
Heisenberg interactions.\cite{santos10} The two independent
choices for the pairing matrix are
$\underline{\Delta}^{(1)}(\mathbf{k})$=$i\Delta_{0}s_{0}\otimes\sigma_{2}$
and
$\underline{\Delta}^{(2)}(\mathbf{k})$=$\Delta_{0}s_{3}\otimes\sigma_{2}$.
For these pairings, it is easy to see that no gap opens in the
topological surface states for both models. Since results for
the two models are identical, we only show those for the model
(I). As shown in Fig. 6 for
$\underline{\Delta}^{(1)}(\mathbf{k})$, a peculiar anisotropic
Andreev bound state structure is observed. An important
difference of this pairing from the above odd-parity pairing
channels is that it is anisotropic with respect to $k_{x}$ and
$k_{y}$. Though the bulk dispersion is gapless in the
$k_{y}k_{z}$ ($k_{x}k_{z}$) plane for
$\underline{\Delta}^{(1)}(\mathbf{k})$
($\underline{\Delta}^{(2)}(\mathbf{k})$), there is still a band
of Andreev bound states for the wave vectors smaller than
$k_{F}$ where a gap opens. The peculiar feature of the Andreev
bound states along $k_{y}$ ($k_{x}$) for
$\underline{\Delta}^{(1)}(\mathbf{k})$
($\underline{\Delta}^{(2)}(\mathbf{k})$) is that they are
dispersion-less (that is, completely flat).

\begin{figure}
\centering
\includegraphics[width=9cm,height=8cm,angle=0]{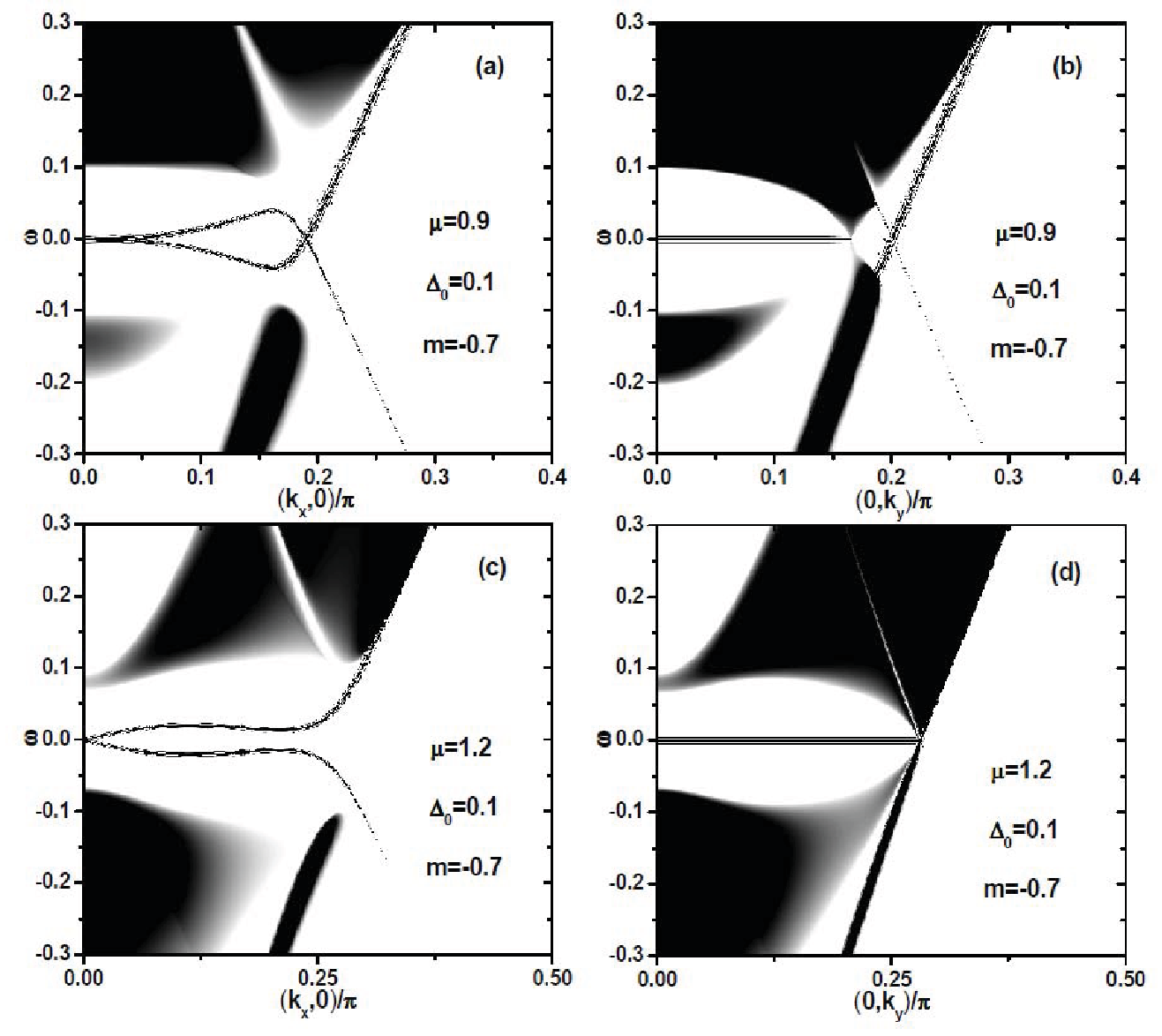}
\caption{Spectral function for odd-parity inter-orbital triplet pairing
in the equal spin pairing channel, for
model (I) and $\underline{\Delta}^{(1)}(\mathbf{k})$=$i\Delta_{0}s_{0}\otimes\sigma_{2}$.
(a) and (c) are along the $k_{x}$ direction while (b) and (d) are
along the $k_{y}$ direction.
The parameters are as shown on the figure.}
\end{figure}

Besides the Andreev bound states within the gap, the
redistribution of spectral weights arising from the particle
hole mixing is also very interesting. An important feature is
the appearance of a linear band beyond the Fermi momentum and
below the chemical potential, existing as a particle hole
symmetric band of the original topological surface states in
the normal phase. Once a bulk superconducting pairing forms in
the topological insulator itself (and not in the intercalated
copper) in Cu$_{x}$Bi$_{2}$Se$_{3}$, such a linear dispersive
band is always there, no matter a gap opens or not in the
topological surface states. To see the above feature more
clearly, we show in Figs. 7(a) and 7(b) the energy distribution
curves (EDC) for several typical wave vectors for two typical
pairings and parameter sets, which are same as those of Figs.
1(a) and 3(a), respectively. The linear band mentioned above
appears as a well defined peak slightly below the chemical
potential. As the wave vector increases and shifts away from
the Fermi momentum, the peak deviates linearly from the
chemical potential and the height and width of it both decrease
rapidly, which is in agreement with the fact that
superconducting pairing forms only close to the chemical
potential. The integrated weight of the linearly dispersive
peaks in Figs. 7(a) and 7(b) are shown in Fig. 7(c), as a
function of the wave vector. If the gap in the superconductors
realized from a topological insulator is larger than what is
reported in Ref. \cite{wray10}, the above linear dispersive
structure in the EDC could be detectable by ARPES for the wave
vectors close enough to the Fermi momentum.\cite{hor10,wray10}
Then this well defined peak structure arising from the
topological surface states could be used as a good indicator of
the formation of superconducting correlation in
Bi$_{2}$Se$_{3}$ and the involvement of the topological surface
states in the superconducting phase.

\begin{figure}
\centering
\includegraphics[width=7cm,height=12cm,angle=0]{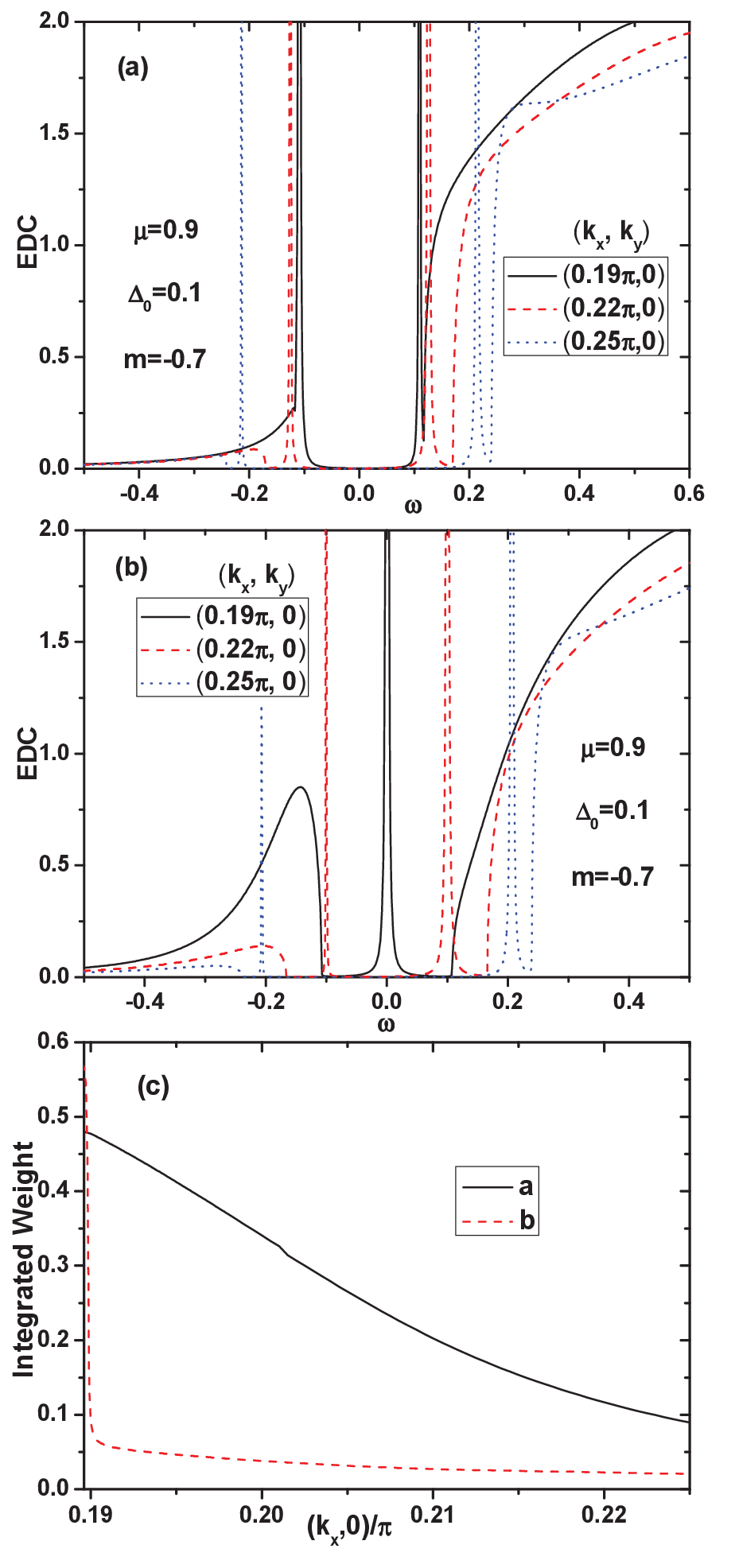}
\caption{EDC for three typical wave vectors for
(a)even-parity intra-orbital singlet pairing
(with parameters same as in Fig. 1(a)) and
(b)odd-parity inter-orbital triplet pairing
(with parameters same as in Fig. 3(a)) within model (I).
The parameters are as shown on the figures.
(c)The integrated weight of the linearly dispersive peaks
slightly below the chemical potential with parameters corresponding to (a) and (b),
respectively.}
\end{figure}

From the above results for five different pairing symmetries,
we observe a simple rule for the existence of nontrivial
surface Andreev bound states. For odd-parity pairings, when a
full gap opens in the continuum part of the surface spectrum
but no gap opens in the topological surface states, a band of
surface Andreev bound states would arise which traverses the
bulk pairing gap. This criterion is verified also by
calculations for other superconducting pairings not presented
here. The results in this subsection are summarized in Table I.

\begin{table}[ht]
\caption{Summary of results for the bulk pairings considered
explicitly in the present work. Results for two models, (I) and
(II), are compared. `TSS' and `ABS' are the abbreviations for
`topological surface states' and `Andreev bound states',
respectively. `$+$' and `$-$' means the even-parity and
odd-parity pairings. For `Gap in TSS', `Y' and `N' represents
that a gap could and could not open in the topological surface
states. For `ABS', `Y' and `N' denotes that Andreev bound
states exist and do not exist on the surface for a certain bulk
pairing.} \centering
\begin{tabular}{c c c c c c}
\hline\hline
$\underline{\Delta}$($\mathbf{k}$) & $is_{2}\otimes\sigma_{0}$
& $is_{2}\otimes\sigma_{1}$ & $s_{1}\otimes\sigma_{2}$
& $is_{2}\otimes\sigma_{3}$ & $is_{0}\otimes\sigma_{2}$ \\ [0.5ex]
\hline
$P$ & $+$ & $+$ & $-$ & $-$ & $-$ \\
Gap in TSS: (I) & Y & N & N & Y & N \\
Gap in TSS: (II) & Y & N & Y & N & N \\
ABS: (I) & N & N & Y & N & Y \\
ABS: (II) & N & N & N & Y & Y \\
\hline
\end{tabular}
\end{table}

\subsection{the surface Andreev bound states}

For some superconducting pairings, such as the $p\pm ip$ wave
pairing, it was known that Majorana fermions exist as gapless
surface or edge modes.\cite{qi10b} According to an argument by
Linder \emph{et al}, all zero energy Andreev bound states
emerging from the nondegenerate (that is, on each surface)
topological surface states should be Majorana
fermions.\cite{linder10} As regards our case, in the parameter
region where the topological surface states are well separated
from the bulk conduction band, one observation is that the
surface Andreev bound states in the gap region connect
continuously to the topological surface states inherited from
the normal phase (e.g., Fig. 3(a)). Since the topological
surface states are spin polarized helical and nondegenerate,
the surface Andreev bound states on each surface should also be
nondegenerate. Then according to the arguments by Linder
\emph{et al}\cite{linder10}, the zero energy Andreev bound
states presented in the above section should also be Majorana
fermions.

To see more clearly the properties of the surface Andreev bound
states, we perform numerical calculations on a finite layer
superconducting film. As an example, we will analyze the
odd-parity inter-orbital triplet pairing channel described by
$\underline{\Delta}(\mathbf{k})$=$\Delta_{0}s_{1}\otimes\sigma_{2}$,
within model (I). Fig. 8(a) shows the dispersion for a fifty
layer film. It reproduces all the basic features in the
spectral function (see Fig. 3(a)). Enlargement of the low
energy dispersion (Fig. 8(b)) shows that dispersion of the
surface Andreev bound states is linear close to the $\Gamma$
point. For all the bulk pairings studied above, the dispersion
for the superconducting film reproduces well the features of
the corresponding spectral function. Figure 9(a) shows the wave
function amplitudes for the surface state localized on the top
several layers. The corresponding behavior for the topological
surface states in the normal phase is presented in Fig. 9(b).
The decay behavior of the surface bound states into the bulk in
Fig. 9(a) is seen to change continuously from oscillatory
exponential decay in the gap region\cite{fu10} to monotonic
exponential decay outside the gap region. That is, the state
changes from particle-hole mixed superconducting quasiparticle
to the topological surface states in the normal phase.

\begin{figure}
\centering
\includegraphics[width=8cm,height=10cm,angle=0]{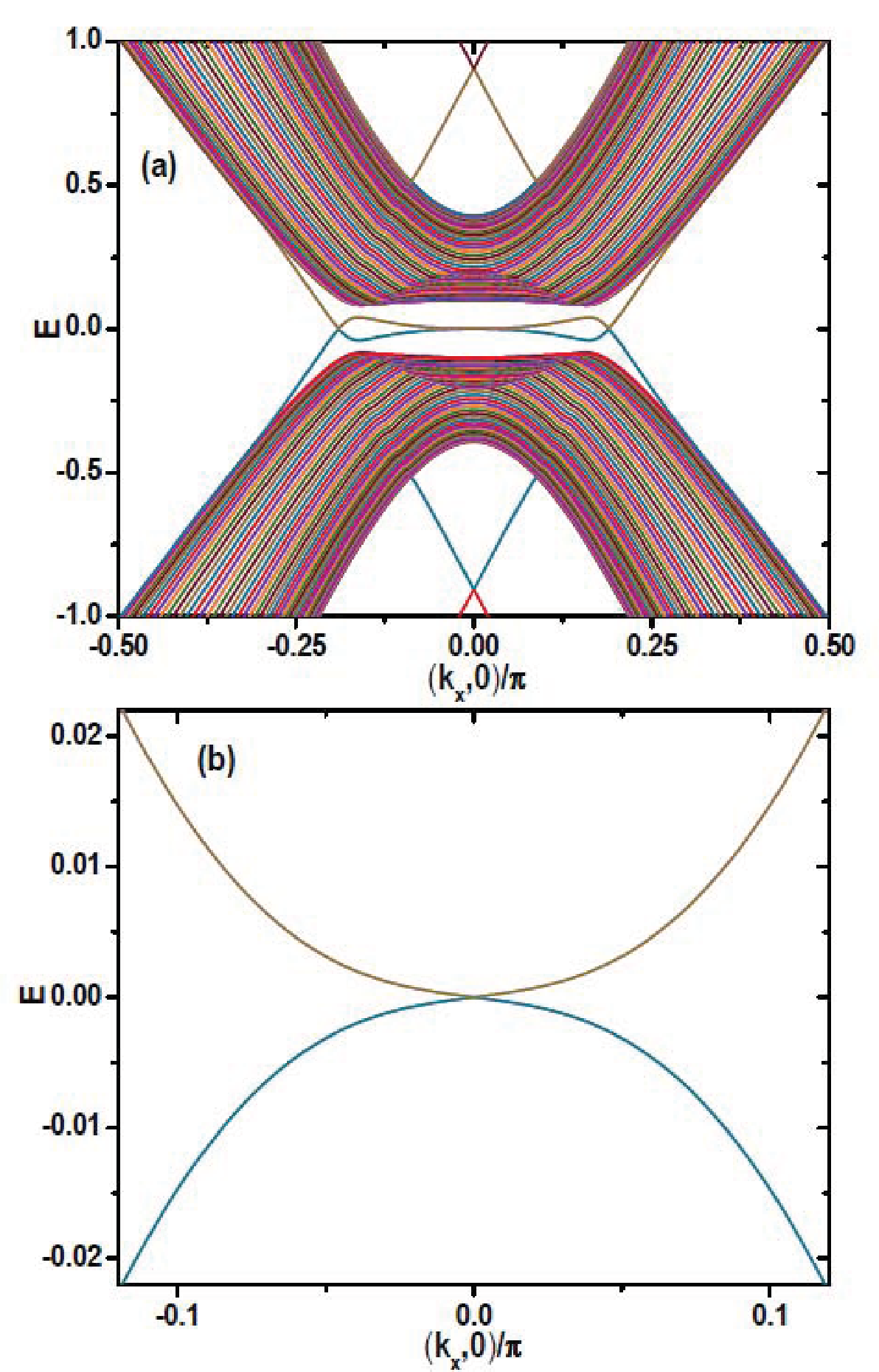}
\caption{(a) Dispersion of a fifty layer superconductor emerging from model (I)
and for the odd-parity inter-orbital triplet pairing $\underline{\Delta}(\mathbf{k})$=$\Delta_{0}s_{1}\otimes\sigma_{2}$.
Parameters used are $\mu$=0.9, $\Delta_{0}$=0.1, and
$m$=-0.7. (b) An enlargement of the small wave vector and low energy part of (a).}
\end{figure}

\begin{figure}
\centering
\includegraphics[width=8cm,height=10cm,angle=0]{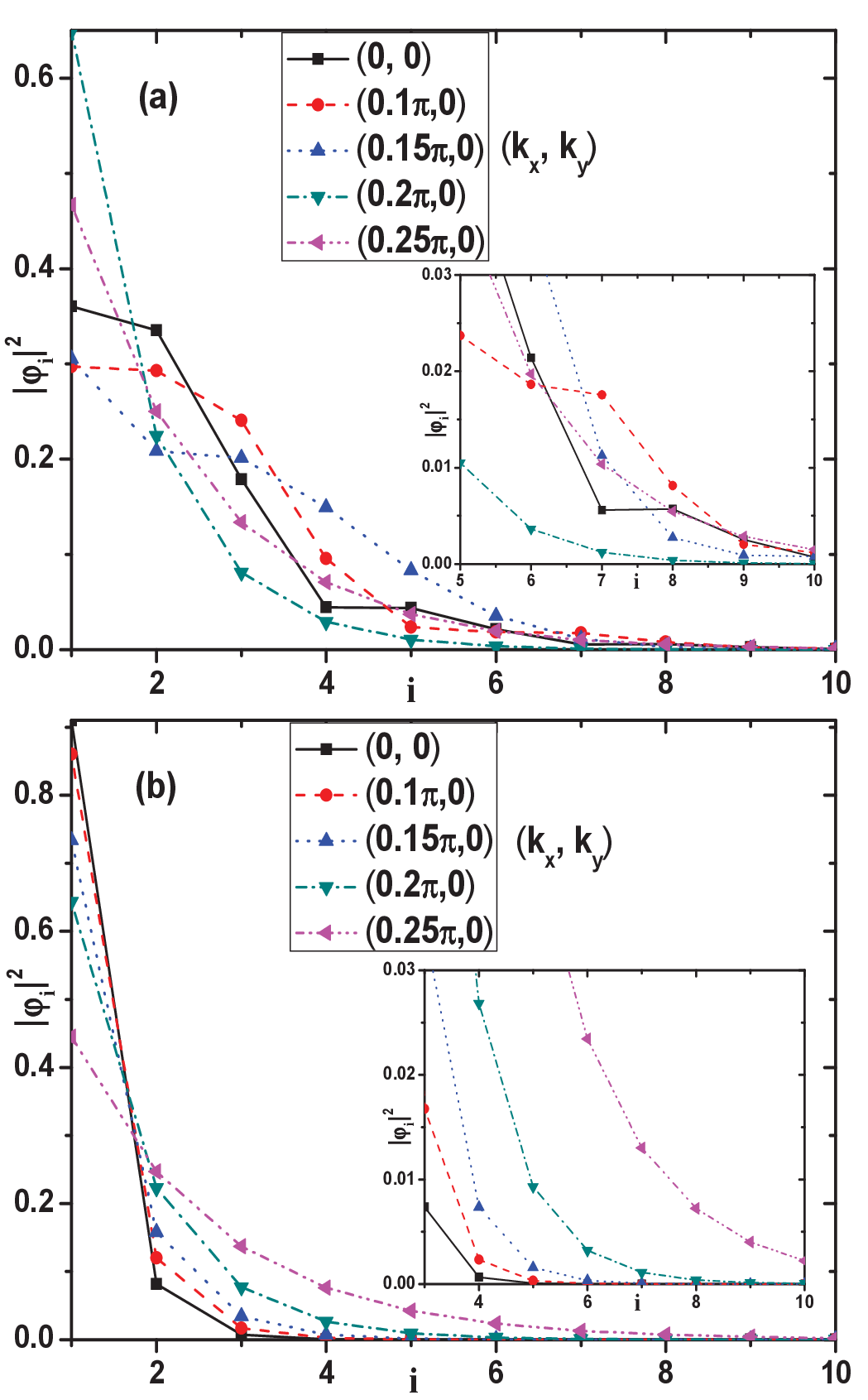}
\caption{(a) Decay of the wave function amplitude with layer number
for the surface Andreev bound states localized on the top several layers,
results are obtained for the superconducting state emerging from model (I)
and for the odd-parity inter-orbital triplet pairing $\underline{\Delta}(\mathbf{k})$=$\Delta_{0}s_{1}\otimes\sigma_{2}$.
The parameters are $\mu$=0.9, $\Delta_{0}$=0.1, and
$m$=-0.7. The Fermi wave vector in the $k_{x}$ direction is
about 0.19$\pi$. (b) Decay of the wave function amplitude with
layer number for the topological surface states localized on
the top several layers, for the normal states of model (I) with
$m$=-0.7. The two insets show enlargements of the small
amplitude regions of the two figures.}
\end{figure}

In this paper, we have always been discussing a homogeneous
phase both in the bulk and on the surface. However, in the
presence of exotic surface excitations as vortices, novel
Majorana fermion modes may appear in the vortex core even for
bulk pairings with no surface Andreev bound states. There have
already been many papers focusing on this possibility, which
usually start from the effective
model.\cite{qi09l,santos10,hosur10}

The Andreev bound states that appear in many different
superconducting pairings as shown above confirm the idea that
superconducting states realized in a topological insulator are
very probable to have nontrivial topological characters. The
Andreev bound states, if exist, should be easily detectable in
a tunneling type experiments as a well defined zero energy
peak. Another way to detect the Majorana fermions as zero
energy Andreev bound states is to take advantage of the various
phase sensitive transport devices proposed to produce and
manipulate the Majorana fermions.\cite{fu0809,law09}

\section{summary}
In this paper, we have discussed the surface spectral function
of superconductors realized from a topological insulator, such
as the copper-intercalated Bi$_{2}$Se$_{3}$. These functions
are calculated by projecting bulk states to the surface for two
different models used previously for the topological insulator.
Dependence of the surface spectra on the symmetry of the bulk
pairing order parameter are discussed with particular emphasis
on the odd-parity pairing. When an odd-parity pairing opens a
full gap in the bulk, but not for the topological surface
states, zero energy Andreev bound states are shown to appear on
the surface. When the topological surface states are well
separated from the bulk conduction band, the redistribution of
spectral weight induced by the onset of superconductivity
produces a linearly dispersive peak structure beyond the Fermi
momentum and below the chemical potential. It is proposed as a
criterion for confirming that superconductivity occurs in the
Bi$_{2}$Se$_{3}$ (and not in copper) and the topological
surface states are involved in the superconducting phase. The
zero energy surface Andreev bound states are argued to be
Majorana fermions.

\begin{acknowledgments}
We thank Peter Thalmeier for helpful discussions. This work was
supported by the NSC Grant No. 98-2112-M-001-017-MY3. Part of
the calculations was performed in the National Center for
High-Performance Computing in Taiwan.
\end{acknowledgments}\index{}


\end{document}